# Transforming India's Agricultural Sector using Ontology-based Tantra Framework

Dr. Shreekanth M Prabhu, Professor, Department of Information Science and Engineering, CMR Institute of Technology, Bengaluru

**ABSTRACT**

Food production is a critical activity in which every nation would like to be self-sufficient. India is one of the largest producers of food grains in the world. In India, nearly 70 percent of rural households still depend on agriculture for their livelihood. Keeping farmers happy is particularly important in India as farmers form a large vote bank which politicians dare not disappoint. At the same time, Governments need to balance the interest of farmers with consumers, intermediaries and society at large. The whole agriculture sector is highly information-intensive. Even with enormous collection of data and statistics from different arms of Government, there continue to be information gaps. In this paper we look at how Tantra Social Information Management Framework can help analyze the agricultural sector and transform the same using a holistic approach. Advantage of Tantra Framework approach is that it looks at societal information as a whole without limiting it to only the sector at hand. Tantra Framework makes use of concepts from Zachman Framework to manage aspects of social information through different perspectives and concepts from Unified Foundational Ontology (UFO) to represent interrelationships between aspects. Further, Tantra Framework interoperates with models such as Balanced Scorecard, Theory of Change and Theory of Separations. Finally, we model Indian Agricultural Sector as a business ecosystem and look at approaches to steer transformation from within.

**Biographical Notes**

Dr. Shreekanth M. Prabhu received his B Tech degree in Metallurgical Engineering in 1984 followed by M Tech degree in Computer Science and Engineering in 1986, both from IIT Bombay and Ph. D from Visvesvaraya Technological University (VTU) in 2020. He worked in IT Industry for 25 years between 1986-2011. His career spanned IT majors such as TCS, IBM and HP as well as a startup, during which he held many leadership positions. Then he joined academia and served PES University, Bengaluru between 2011 and 2019. His research interests include Information Management, E-Governance, Social Networks, AI and Machine Learning as well as applying technology to better understand Economics and Linguistics. He has also written many articles in the field of Governance, Strategy and Education. He is currently working as Professor, Department of Information Science and Engineering, CMR Institute of Technology, Bengaluru, India.

**Acknowledgements**: Author acknowledges the valuable feedback on this research received from Dr S. G. Deshmukh, Professor, Department of Mechanical Engineering, IIT Delhi and Dr K. N. B. Murthy, former Vice-Chancellor, PES University, Bengaluru and Dr. Natarjan S, Professor, Department of Computer Science and Engineering, PES University, Bengaluru

# 1. Introduction

India's agriculture sector is one of the largest in the world. "In 2017-18, India produced 275 Million Tons of food grains. India is the largest producer (25% of global production), consumer (27% of world consumption) and importer (14%) of pulses in the world. India's annual milk production was 165 MT (2017-18), making India the largest producer of milk with world's second-largest cattle population (190 million in 2012). It is the second-largest producer of rice, wheat, sugarcane, cotton and groundnuts, as well as the second-largest fruit and vegetable producer, accounting for 10.9% and 8.6% of the world fruit and vegetable production, respectively" (FAO, 2019).

However, Indian agriculture sector compared to other countries involves far too many people with low per-capita contribution to GDP. In the last few years capital formation in agricultural sector has declined from 15 percent to 13 percent. Today only 23 percent of rural GDP is agricultural, and increased numbers are taking up non-farm jobs. Average farm holdings also have been reduced from 2 hectares to 1 hectare. Public investment in agriculture is stagnant. Agriculture wages also have declined. From time-to-time Governments have announced loan waivers to agricultural sector incurring huge costs to exchequer. The loan waivers have not had major impact on incomes of farmers nor on the health of the sector as a whole. Majority of institutional credit goes to large farmers who benefit due to loan waivers while small and medium farmers continue to borrow from money lenders at exorbitant rates of interest. There continue to be cases of distress in agricultural sector due to crop losses, income losses (due to glut in the market) and indebtedness. Government's policies in favour of cereals also have resulted in distortions in cropping patterns and led to producer orientation instead of consumer orientation. Farm markets needs to be freed from excessive state intervention(The Hindu Podcast, May 2019)

In general, Government of India has responded with complex and confounding policy interventions to support the agricultural sector. Most recently to address the continued distress in farm sector Government of India launched PM-Kisan scheme to provide basic annual income to farmers with huge outlay. Table 1 below gives details on budgetary allocations Government of India made to different schemes in the last 3years (Mathew, 2019).

It is observed that in many cases allocations are far less than outlays described in scheme descriptions. Governments have a fiscal challenge to fund all these schemes. Government also provides subsidies related to power and fertilizer, which are not included in Table 1. The above are just allocations of the Central Government. Over and above this, State Governments have their own allocations and schemes. Despite all these allocations, Alagh opines that many schemes have last mile issues and do not reach farmers (Alagh, 2018).

It is apparent that Government of India plays an all-pervasive role in the agricultural sector. Above all, Government tightly regulates agricultural markets and exports. There are laws that restrict sale of farm land for use other than farming. There are also several cases where farmers commit suicide due to variety of reasons. Most often indebtedness to money-lenders is the major cause. Here again Government compensates the families.

Current situation in agricultural sector, calls for advanced information system that can provide comprehensive analysis of Indian agricultural sector that enables Government to design/choose right interventions at right funding levels as well as formulate strategies so that sector needs less and less Government support and intervention.

In this paper we approach the analysis of Indian agricultural sector from three different view-points: descriptive, normative and transformative. When we use descriptive viewpoint, the objective is to describe sectoral and societal information in a unified manner as consistently as possible. With the normative viewpoint, the objective is to design the information system so that it can help achieve a set of goals for the sector. With transformative view-point, the idea is to use the business ecosystem paradigm to take the sector to next stage of evolution. Here ecosystem participants compete as well as cooperate. They act as agents who may self-organize. They evolve on their own as well as coevolve. They adapt to the environment as well as collectively become part of emergent phenomena.

**Table 1: Agricultural Allocations (in Rs. Crores)**

| Sr.No | SCHEME | Nature of Support | 2018-19 | 2019-20 |
|---|---|---|---|---|
| 1 | Pradhan Mantri Kisan Samman Yojana (PM-KISAN) | Rs 6,000 per year is provided as Minimum Income Support to Farmers | 20,000 | 75,000 |
| 2 | Interest Subsidy for short-term credit to Farmers | Interest subvention (2% on grant of loan and 3% additional on repayment). The loan has INR 3,00,000 limit. | 14,987 | 18,000 |
| 3 | Pradhan Mantri Fasal Bima Yojana | Crop Loss Insurance. Covers yield-losses, prevention of harvesting due to weather, post-harvest losses and local calamities at individual farm level. | 12,976 | 14,000 |
| 4 | Rastriya Krishi Vikas Yojana | Central Grants to State Governments to invest in agriculture | 3,600 | 3,745 |
| 5 | Pradhan Mantri Krishi Sinchai Yojana | Extend cover of irrigation and increase the efficiency of water usage i.e. more crop per drop. | 2,955 | 3,500 |
| 6 | Market Intervention Scheme and Price Support Scheme | MIS takes care of horticultural products and PSS takes care of Minimum Support Price (MSP) for 22 commodities | 2,000 | 3,000 |
| 7 | National Mission for Horticulture | Center and State contribute at 60:40 ratio for integrated development of horticulture | 2,100 | 2,226 |
| 8 | National Food Security Mission | Food grains (rice and wheat) provided at very low rates to nearly 66% of population. Outlay is in Lakhs of Crores). | 1,510 | 2,000 |
| 9 | Pradhan Mantri Annadata Aay SanraksHan Yojana | PM-AASHA is an umbrella scheme that combines 3 schemes: Price-Support (PPS), Price-deficiency Payment Scheme (PDPS) and Pilot Private Procurement and Stockist Scheme (PPPS). | 1,400 | 1,500 |
| 10 | Integrated Scheme on Agricultural Marketing | This includes Marketing infrastructure including storage, support for integration of value chain and electronic marketing. | 500 | 600 |
|  | Total |  | 67,800 | 1,30,485 |

Rest of the paper is as follows: **Section 2** covers **Analysis of Indian Agricultural Sector**. **Section 3** covers **Ontology-based Tantra Social Information Management Framework. Section 4, Modeling and Analysis of Agricultural Sector using Tantra Framework** describes how Tantra Framework can be applied to Agricultural Sector and operationalized. **Section 5, Transforming the Agriculture Sector**, covers how the mutual dependency between the sector and Government can be radically reduced. Finally, in **Section 6,** we provide concluding remarks.

## 2. Analysis of Indian Agricultural Sector

In this section we analyze the prevailing issues in Indian Agriculture Sector and reforms under progress. This will act as a back-drop to proposed use of Tantra Information Framework in succeeding sections.

According to Puja Mondal, ten major problems confronting Indian Agriculture are: Small and fragmented land holdings; Seeds; Manures, Fertilizers and Biocides; Irrigation; Lack of Mechanization; Soil Erosion; Agricultural Marketing; Inadequate Storage Facilities; Inadequate Transport and Scarcity of Capital. The solutions suggested by her are described in Table 2.

Jagannathan (2018) in his article reiterates some of the issues above. Further he talks about low per-capita contribution of agricultural sector to the national GDP: agriculture's share of gross domestic product (GDP) is one-third the size of the population dependent on it. This may need increase in non-farm jobs so that fewer people are dependent on agriculture. Second issue he raises is the obsession with subsidies in place of investment in cold-chains, warehousing, irrigation and inter-connection of markets. According to him Rs 16,300 per hectare is paid out to support agriculture for seeds, diesel and procurement bonus. This is over and above 2.61 lakh crore subsidy related to food and fertilizer. Other issues he raises is restriction on free flow of goods in markets, overly centralized policies on deciding prices. Then he suggests that farmers are encouraged to do dairy farming and explore contract farming, etc. He also advocates easier processes to change land-use from agricultural to commercial/residential.

Governments have to make hard policy choices while managing agricultural sector. In a recent article, Gulati (2020) deplores the bias in favour of consumer and locking of resources to fund the food security program. Government spends to the tune of Rs 1,55, 570 crores to support Food Security Program. In 2013, Government of India dramatically increased the coverage of subsidized food grains to 66% of populace, this in turn may have hurt the growth in agricultural sector. This was done despite the wiser counsel of Planning Commission. But real cost according to Gulati is much higher as the procurement agency, Food Corporation of India has to borrow from myriad sources to procure the huge amounts of grains.

Last few years Government has managed to kept the inflation low. This in turn has depressed the profitability in agricultural sector. Swift policy actions also have meant even with high price volatility, neither farmers nor hoarders can earn high profits for too long. Table 3 lists some more issues confronting the agricultural sector. Table 4 lists some of the ongoing reforms as well as some suggested approaches to address the issues in agricultural sector.

**Table 2: Ten Major Problems in Indian Agricultural Sector and their Solutions**

| Sr. No. | Problem | Suggested Solution |
|---|---|---|
| 1 | Small and fragmented land holdings | Consolidation/Cooperative Farming |
| 2 | Distribution of high-quality seeds at affordable prices | Government may need to support small and marginal farmers. |
| 3 | Manures, Fertilizers and Biocides increase productivity but cause soil erosion | Need to balance the use and adopt natural farming |
| 4 | Irrigation has helped certain states enormously while many others depend only on Monsoon | Irrigation needs to be expanded but its ill-effects should be kept in mind |
| 5 | Lack of Mechanization in farming methods | Training and support to farmers |

| 6  | Soil Erosion | Treatment of soil to make it fertile |
| 7  | Agricultural Marketing: Farmers continue to be short-changed. | Reforms should continue here |
| 8  | Inadequate Storage Facilities | Providing Rural Storage Centers |
| 9  | Inadequate Transport | Rural Roads are being expanded |
| 10 | Scarcity of Capital | Now major portion of credit is institutional but still money lenders persist |

**Table 3: Issues in Agricultural Sector**

| Sr. No | Issue | Remarks |
|---|---|---|
| 1 | Water Planning | Water is not equitably distributed regionally. Water tables are depleting everywhere. We have states like Punjab which are water deficit growing crops such as Rice. Many farmers grow Sugarcane which also takes lot of water for commercial reasons (Jebaraj,2018). Even with huge investment in irrigation outcome was suboptimal in many cases due to last mile issues. |
| 2 | Land | Land is a scarce resource which is needed for industrialization as well as agriculture. While ideally farmers should be free to sell their land to commercial use, excessive urbanization will weaken rural economies. |
| 3 | Environment and Farming Practices | Periodic crop losses due to ill-timed rains, floods and drought put farmers in distress. Use of fertilizers in a non-judicious manner leads to erosion of soil. Pesticides increase the cost as well as spoil the health. Cheap power has meant over-drawing of water and depletion of water tables. |
| 4 | Economy | Agriculture engages nearly half the population but accounts for less than 20 percent of GDP. The combined loan waivers announced by 4 states in 2017 amounted to USD 13.6 billion. |
| 5 | Governance | Many Government schemes such as Minimum Support Price helps rich farmer lot more than their poorer counterparts. Typically, rich farmers purchase from smaller farmers and sell to the Government. In general targeting of schemes remains a weak point. |
| 6 | Supply chain and regulation | Farmers are restricted to sell in their local Mandis leading to inefficient supply chain. Studies have shown that a farm produce sometimes changes ten times without making any value-addition. Both producers and consumers feel short-changed. |
| 7 | Market Distortions | As Governments buy huge quantities of rice and wheat it affects cropping patterns. A year of drought may be followed by bumper harvest. In either case farmers and/Government will have to bear the burden of supply shocks. |

| 8 | Market Separations | Farmers in many cases are forced to sell in the local Mandis. This itself is a barrier. Due to variety of reasons, farmers may be forced to sell to aggregators. |
| --- | --- | --- |
| 9 | Storage and distribution | To support farmers, Governments, procure lot more grains than the actual need. This needs huge storage. Lack of adequate storage means grains are stored in the open and lead to rotting of grains. Currently only 10 percent of requirement is met as far as storing wheat in steel silos is concerned |
| 10 | Trading | As a part of the procurement function, there is need for people/institutions who do quality assessment of the crops. Traders would also like to avail of insurance facilities when they are trading with unfamiliar parties/locations. |
| 11 | Lack of quality information | It is much easier to target schemes to farm owners than tenant farmers or farm workers due to lack of information. As a result, those who actually need a benefit may not get it. |
| 12 | Return on investment | The profitability in crop cultivation from public irrigation hardly matches with the opportunity cost of public irrigation. Public irrigation projects are very capital intensive but the financial outcomes hardly make justice to investment. |
| 13 | Lack of quality control | In many cases there is inadequate quality control of grains as it moves through the supply chain. At times inferior grains are purchased and at other times their quality declines as they move through the supply chain. Quality is a key issue to be addressed to liberalize the sector. |

**Table 4: Reforms in Agricultural Sector**

| Sr. No. | Reform | Remarks |
| --- | --- | --- |
| 1 | Electronic Nationwide Agricultural Market (ENAM) | This is an important initiative. This enables nation-wide trading of agricultural goods, compared to the prevailing Mandi system where a farmer has to sell only in the local Mandi. ENAM requires traders to pay license fees to be able to trade in the market. Government needs to assure trust, quality as well as safe and secure transit of goods. |
| 2 | Organic Farming | Organic farming fetches higher prices and it is environment friendly. Up-front investment in organic farming is far less and depending on the yield it may even fetch better returns. Sikkim has declared itself to be first 100% organic state in India and won global recognition. There is lot of export potential for organic produce. |
| 2 | Zero-budget Natural Farming (ZBNF) | This is revival of traditional Indian farming technique which relies of cow dung and vegetable residue on account of bio-diversity in farms. ZBNF(Swarajya Videos, 2019) is lot more environment friendly even compared to organic farming and it does not involve buying organic compost from any of the companies. |

| 3. | Reducing Centralizing procurement | The State Governments that run Public Distribution System (PDS) can procure food grain directly rather than get it through the Food Corporation of India (FCI). The role of FCI can be just confined to maintaining the stocks that can help overcome any food emergency. So far, this proposal has not got traction. |
|---|---|---|
| 4 | Reforming Agricultural Practices | Farmers can be incentivized to shift to better farming practices by giving better health and education facilities. Good practices also include water conserving practices in agriculture. |
| 5 | Support Rural Economy | Agriculture sector cannot sustain itself in villages if villagers steadily leave for cities and towns. It is important to make use of E-commerce to distribute growth instead of blind pursuit of urbanization. |
| 6 | Imaginative solutions to financial distress | Politicians need to move from price support policies or loan waivers to income/investment support on a per acre basis. The payment can happen using Direct-Benefit-transfer schemes. This can be both good politics and economics. |
| 7 | Reform PDS and Food Security regime | Government should move to Flexible Social Benefit Plans (Prabhu, 2015) and optimize the amounts of food grains produced and target them to those who really need them. In some backward areas broad-brush approaches may have worked. But as districts become more prosperous, people should be given greater choice. |
| 8 | Use of Science and Technology | Use of Biofortified food can move India from food security to nutrition security. Government also needs to carefully evaluate introduction of Genetically Modified Foods. Use of Solar Energy, Wind Energy in farming as well as adopting drip irrigation methods. |
| 9 | Doing away with income tax exemption for Agricultura income | The tax exemption helps only rich farmers. On top of that many use it to declare non-agricultural income as agricultural income and evade tax. This provision disempowers agricultural sector to be self-sustaining. On top of that it drains resources from economy at large to support agricultural sector. |

## 3. Ontology-based Tantra Social Information Management Framework

The name *Tantra Framework* is chosen to reflect enormous connectivity of information through the framework. Etymologically in Sanskrit "tantram", literally means "loom, warp," from "tan", "to stretch, extend", hence, figuratively, "***groundwork, system, doctrine***".

The objective of Tantra Framework is to represent social information in a unified manner covering all aspects, linkages between aspects while accommodating different levels of detail using a process called reification. Tantra Framework extends Zachman (2003,2007) Framework by adding three additional columns namely relators, relationships and separations to the six interrogative columns in Zachman Framework. Zachman treats his framework as Ontology. In Tantra Framework the ontological approach is further strengthened as the concept of relator and relationship are taken from Unified Foundational Ontology (Santos Jr., 2013). The framework thus is well-equipped to manage knowledge pertaining to any domain.

The aspects of Tantra Framework useful to express social information in a comprehensive manner are given below.

1. People/Communities/Categories (Who)
2. Places/Addresses/Locations/Zones (Where)
3. Assets/Attributes (What)
4. Events (When)
5. Processes (for enrolment, intervention) (How)
6. Metrics to measure development (Why)
7. Relationships (between aspects)
8. Relators (enable relationships)
9. Separations (difficulty of establishing a relationship)

The Table 5 below, gives the view of Tantra Framework. Each aspect in Tantra Framework is reified using perspectives at contextual(named), conceptual(defined). Logical(designed), physical(configured) and instantiated levels. Table 6 and 7 describe how different aspects get reified.

**Table 5: Tantra Framework**

| Perspectives | Aspects | | | | | | | | |
|---|---|---|---|---|---|---|---|---|---|
| | Who | Where | What | When | How | Why | Relationships | Relators | Separations |
| **Contextual (Named & Scoped)** | Name | Name | Name | Name | Name | Name | Name | Name | Name |
| **Conceptual (Defined)** | Concept | Concept | Concept | Concept | Concept | Concept | Concept | Concept | Concept |
| **Logically Designed** | Relation | Relation | Relation | Relation | Relation | Relation | Relation | Relation | Relation |
| **Physically Configured (Schema)** | Network schema | Network schema | Network schema | Network schema | Network schema | Network schema | Network schema | Network schema | Network schema |
| **Detailed/Instantiated** | Unique ID | Unique ID | Unique ID | Unique ID | Unique ID | Unique ID | Unique ID | Unique ID | Unique ID |

**Table 6: Reification of People Domain**

| Perspective | All people | Citizens | Residents | Resident Aliens | Resident Citizens |
|---|---|---|---|---|---|
| **Named (Identified & Contextualized)** | All the people known and to be known to the framework. | People who are citizens | People who are residents | People who are resident but alien | Resident Citizens |
| **Defined (Conceptually Structured)** | What makes one a member of this domain/role | What makes one a member of this domain/role | What makes one a member of this domain/role | What makes one a member of this domain/role | What makes one a member of this domain/role |
| **Logically Designed** | Related attributes that map to other aspects. | Related attributes that map to other aspects | Related attributes that map to other aspects. | Related attributes that map to other aspects. | Related attributes that map to other aspects. |
| **Configured** | Representation in Graph database as nodes and edges. | Representation in Graph database as nodes and edges. | Representation in Graph database as nodes and edges. | Representation in Graph database as nodes and edges. | Representation in Graph database as nodes and edges. |
| **Instantiated** | Instantiate with unique ID. | Instantiate with unique ID. | Instantiate with unique ID. | Instantiate with unique ID. | Instantiate with unique ID. |

Table 7: Reification of other aspects

| Aspect | Examples |
|---|---|
| **Address** | Residential address, General Address/Location, Commercial Address, Institutional address, Address for civic amenity |
| **Event** | Birth event, Education Enrolment, Employment, Marriage, Retirement, Death, etc. |
| **Asset** | House Owned, Vehicle Owned, Land Owned, Business Owned, Stocks, etc. |
| **Processes** | Aadhaar Enrolment, Voter Enrolment, PDS Enrolment, Birth Registration, Property Registration/Lease |
| **Artifacts** | Aadhar Id, Voter ID, Public Distribution Service (PDS) card, Permanent Account Number (PAN) Card, Birth Certificate, Death Certificate, Property Sale/Purchase, Property Transfer/Ownership. Lease |
| **Relator** | Enrolment Agencies and Service Providers. |

The strength of Tantra Framework is its generic approach. Tantra Framework is implemented using Neo4J graph database which supports flexible yet powerful modelling and visualization.

An entropy construct was defined as an instrument of validation for Tantra Framework (Prabhu,2019). Tantra Framework has been applied to variety of scenarios (Prabhu et, al, 2018 and, 2019). This work is off-shoot of Ph.D. Thesis (Prabhu, 2019)

## 3.1 Tantra Framework as Normative Framework

When Tantra Framework enables representation of information in a comprehensive manner, it only exploits descriptive power of the framework. If we want change to happen, we need to make use of the framework in a normative mode. To facilitate such an exercise, we define Tantra Normative Framework in Figure 5 below. Here, the process of change comprises of setting goals, designing interventions and measuring separations (which in turn leads to recalibrating the goals and interventions as things change) operates on information space created using Tantra Framework.

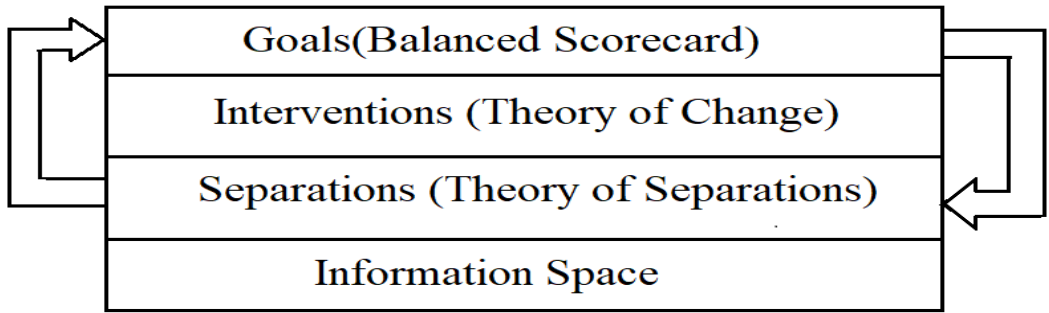

**Figure 1: Tantra Normative Framework**

In Figure 1, above,

1. To set goals, we propose use of ideas from Balanced Scorecard Framework (Kalpan,2010) to cover customer, financial, process, strategic and ethical dimensions while setting goals. The objective column of Tantra Framework enables setting appropriate goals and tracking of related measures. Here consumer, farmer, farm workers, traders and other intermediaries and institutions/individuals involved in financing/investing in the sector can all be considered as customers. We can incorporate goals of Government of India such as doubling farmers' income here.
2. To realize goals, Governments need to design and deploy interventions. These may be schemes such as PM-Kisan or ENAM. To ensure the right policy interventions are chosen we propose use of the paradigm of theory of change (Weiss, 1995). Here an intervention is evaluated in a comprehensive manner by looking at all assumptions and linkages working backwards from outcome to be achieved. The process column of Tantra Framework enables definition of intervention processes that includes inputs, tasks, outputs, outcomes and relationship to other processes. Tantra Framework facilitates continued monitoring of interventions.
3. To assess the state of agricultural sector we use of Bartels' (1968) theory of market separations. Bartels identified financial, spatial, temporal and informational separations as inhibitors that come in the way of healthy functioning of a market. Bartels' theory is applicable to particularly suited to Agricultural Sector.

In the context of this paper, Agriculture and Farmer Welfare Ministry as well as other related ministries can set goals. Bodies such as NITI Aayog can design appropriate interventions. Tanta Framework can be used to assess the actual outcomes on the ground and identify process bottlenecks if any. Funding distribution among different schemes is an important lever for Government to manage the effectiveness of spending.

## 4. Modelling and Analysis of Agricultural Sector using Tantra Framework

Indian Agricultural Sector is probably one of the most humongous sectors affecting the lives of a billion people in one way or another. Figure 2 below depicts the Indian Agriculture sector, with all the players and interfaces involved in a generic manner. This is followed by Table 8 that describes the characteristics of sector participants and institutions.

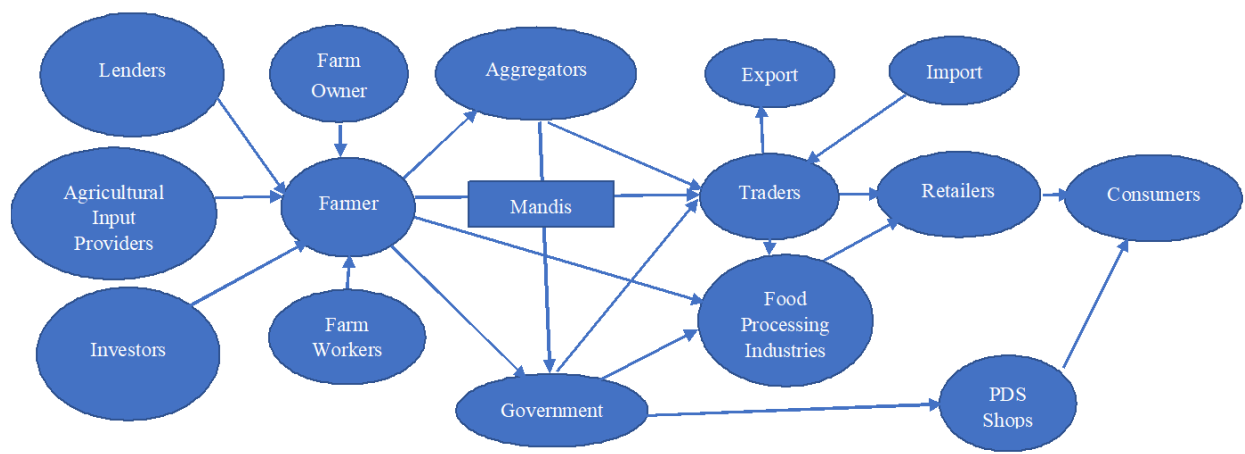

**Figure 2: Indian Agricultural Sector**

**Table 8: Participants in Indian Agricultural Sector**

| Lenders | These are banks, micro-finance companies and local money lenders. Local money lenders occupy a significant portion. There are specialized banks such as NABARD. |
|---|---|
| Investors | A farmer may invest his own money or take from other investors. Contract Farming companies and Food Processing companies invest in agriculture and then buy the produce from farmers. |
| Agricultural Input Providers | Farming requires good quality seeds, soil and water. Modern farming relies heavily on fertilizers and pesticides. Farmers in turn depend on power at low rates for watering their fields. |
| Farm Owners | Farm owners may choose to do farming on their own or give it to a tenant Farmer to do the farming. As a part of land reform large amounts of land was transferred from farm owners to tenant farmers by Government, disregarding property rights. About 85% of farmers now have small holdings. |
| Farmers | Farmers may be Farm Owners or Tenant Farmers. Farm owner may involve his own family in farming and/or engage Farm workers. |
| Farm workers | Farm workers work on farms. They do not necessarily get benefit if price of produce increases |
| Farmer Co-operatives | In case of certain crops such as sugar, farmers have established co-operatives. The idea is to give fair price to farmers. These cooperatives are very popular in states like Maharashtra, Punjab, Karnataka etc. |
| Mandis | Farmers typically sell their produce in Government approved Mandis (Markets) to traders at those Mandis. In case of certain produce, farmers are forced to sell at Mandis by law. |
| Traders | Traders buy from Farmers at Mandis or buy directly from farmers and then sell at Mandis by aggregating produce. Only authorized persons can sell or buy at Mandis. |
| Retailers | The retailers buy from wholesale traders and fulfill requirements in neighborhood. Retailers may also be large stores which are in multi-brand retail. |
| Exporters | Exporters focus on exporting items in demand in export markets. At times exporters are restricted from exporting when local production falls below and prices increase. For example, Basmati Rice to vegetables and fruits. |
| Importers | Certain agricultural produce is imported in large quantities when local production cannot meet the demand. For example, palm oil is generally imported from South-East Asia. |
| Consumers | Consumers may belong to wide range of economic categories and may purchase from open market or rely on public distribution system. |
| Government | Government plays an all-pervasive role in the agricultural ecosystem. For example, there are several arms of Government which decides on Support Prices of 20-30 odd products. Agriculture costs and Prices commission under Ministry of Agriculture recommends MSP for 24 crops. |
| Regulators | APMC act regulates trading and farmers have to sell at designated Mandis. E-NAM which is newly introduced will also need to be regulated. |

Table 9 provides details on the role of Government in Agricultural Sector, in a generic manner. Government has also set up many institutions to serve the agricultural sector such as National Bank for Agriculture and Rural Development (NABARD).

**Table 9: Financial Support provided by Government of India**

| Government Support to Farmers | Description |
|---|---|
| Input Support Means support | Subsidized Fertilizers in particular Urea provided to farmers. The subsidy money however goes to the Fertilizer companies |
| | Subsidized/Free Electricity. Here State or Central Governments bear the cost of power |
| | Seeds at reduced prices to Farmers |
| | Government also supports installation of solar pumps and the like in farms. |
| Price Support | The price support is in the form of Minimum Support Price mainly to rice and wheat. There are 23 crops Government tracks and more and more crops are included in the minimum support price scheme. These Prices may be higher than market at times. Madhya Pradesh Government has come up with an alternative scheme called Price Deficiency where Government instead of procuring grains pays the difference between Market Price and Average Sale Price (ASP). |
| Income Support | Government should strive to ensure market prices to be high enough to ensure desired income to farmers through appropriate price support. There are two schemes for income support: (i) A2+FL +x% factors all input costs farmer directly bears and cost of family labour (ii) C2+x% factors all costs included in A2+FL as well as imputed costs for land and interest on capital. Government generally uses the first scheme. |

| Loan Support | Crop loans are given at a lower rate where Government picks up part of the interest using interest subvention schemes. Many cases loans are waived. UPA Government waived Rs. 71,000 Crores in 2009. |
|---|---|
| Risk Management | Government pays most of the premium for crop insurance. Government pays when crops are damaged or lost, directly or through insurance schemes. |
| Distress/Disaster Support | Government will intervene as appropriate in case of floods, droughts and other emergencies. |
| Investment Support | Telangana Government has Raitabandhu scheme where Government pays cash to farmers depending on acres he tills/owns, which they can use to buy seeds, fertilizers and pesticides. |
| Cash Support | Recently Government of India announced Rs 6,000 annual payout to poor farmers. This is going to cost Government 75,000 Crores benefitting 14.5 Crore farmers. |
| Income tax relief | Any income earned from proceeds of agricultural activities is exempt from income tax. |
| Mahatma Gandhi National Rural Employee Guarantee Act | This provides work opportunities for up to 100 days in Rural Sector. Even though anybody can make use of it. This is particularly helpful to farm workers during lean season. |
| Public Distribution System and Food Security Act | Government buys Food Grains at MSP and then provides to people at very low prices. In case of BPL (Below Poverty Line) category, it is almost free. In case of APL (Above Poverty Line) category they have no pay nominal rate. Up to 30 Kgs of food grains are made available to a family. Under Food Security act, the coverage is increased to 66 percent of population. |
| Government's intervention in Agricultural Market | Government disallows free flow of agricultural -goods. Both export and import are regulated. Government sets Minimum Support Prices which are at times more than market prices. Government's support may mean farmers do not focus enough on productivity of farming. This makes sector less competitive. In addition, MSPs direct crop choices thus causing deficits of items like pulses, fruits and vegetables and excess of cereals or any item that Government supports. In addition, giving fertilizer subsidy means many inefficient producers continue to survive. Farmers have less incentive to move to organic and sustainable agricultural practices. |

Tantra Framework can be used to model agricultural sector along with the overlapping societal dimension. Table 11 below maps aspects of Tantra Framework to the participants and entities in the agricultural sector.

**Table 10: Modelling Agricultural Sector using Aspects of Tantra Framework**

| Aspect | Participants/Entities Modelled |
|---|---|
| **People (Who)** | Farm-owners, Owner Farmers, Lease-holders to Farm land, Tenant Farmers, Marginal Farmers, Small Farmers, Rich Farmers, Aggregators, Traders, Retailers, Consumers, PDS Beneficiaries, APL/BPL beneficiaries., Money Lenders |
| | Farmer Co-operatives, Communes, Trader Associations, Consumer Groups, |
| **Plots (Where)** | Farm Plots, Vacant Plots, Uncultivated Plots, Residential Plots, Commercial Plots, Industrial Plots, Forest Land, Lakes and water bodies, Districts, States, Regions (Forward/Backward), Areas (Irrigated/Prone to Drought) |
| **Events (When)** | This may be announcements on schemes, payment of benefits, procurement or any other event such as transfer of ownership of land. Life cycle events of farmers such as death and migration can also be linked and tracked. |
| **Assets (What)** | Own house, Vehicle, Tractor, Solar Pump, Land Holding, Dairy, Business Owned, Equity, Gold, |
| **Crops (What)** | Rice, Wheat, Sugarcane, Pulses, Vegetables, etc. |
| **Benefits received from Government (What)** | Input Support, Price Support, Income Support, Insurance Support, Interest Subvention, Support, Compensation, Loan Waived (In case of farmers), Fertilizer Subsidy, Electricity Subsidy, Water Subsidy |
| | PDS, MGNREGA, Subsidies related to health, education, electricity, water, LPG and many others |
| **Incomes (What)** | Income through Farming, Income through Farm Lease, Farm Labour Income, Cottage Industry, Animal Husbandry, Non-Farm Income, Money Lending |
| **Processes (How)** | These may be operational processes/interventions to implement Government Schemes or Intervention Process using Theory of Change Methodology (Steins and Valters, 2012) is described in Table 11 below. |
| **Farming Process** | Modern Farming, Organic Farming, Natural Farming |
| **Measures (Why)** | GDP (Agriculture, per-Capita Farmer), % Export, debt level, poverty level, unemployment level, profitability level |
| **Relator** | Banks, Government Agencies, PDS, Insurance Firms, Seed Providers, Fertilizer Firms, etc. |
| **Relationships** | Tenancy, Under Organic Farming, Natural Farming, meaningful relationship between any two aspects. |
| **Separations** | Detailed in Table 12 below. |

**Table 11: Theory of Change Methodology applied to Price-Deficiency Support Intervention**

|  | Price-Deficiency Support instead of MSP |
|---|---|
| **Summary Statement** | Compensating farmers when average price falls below Minimum Support Price (MSP) level by paying them difference between Average Market Price and MSP when they sell in the local market. |
| **Problem Statement** | Farmers have to wait with food stock till Government buys at MSP leading to holding cost and spoilage. Alternatively selling in the market means they do not get as good a price as MSP. |
| **Overall Goal** | Enabling farmers to sell quickly and realize reasonable returns. Reduce the financial burden on Government it has to incur in the procurement and storage. |
| **Change Process** | Encouraging farmers to sell locally and stand up on their own feet. |
| **Change Markers** | % of farmers taking advantage of price-deficiency scheme. |
| **Meta-Theory** | Time is of essence and even a good scheme is useful only if it is timely |
| **Inputs** | Appropriate communication to farmers and traders using variety of media. Spreading the news through communities. |
| **Actors** | Farmers, Government Officials, Traders all play a role. Those who have vested interests to keep the prices low may act as spoilers |
| **Domains of Change** | Marketing and Environment are two important domains of change. When environment is good, crop production may be high leading to fall in prices. Then this scheme should protect interest of farmers. When environment is bad and if it results in crop loss then such loss should be manageable. |
| **Internal Risks** | If average market price is manipulated then it can lead to Government making excessive payments. Thus, Price Deficiency Support Payment as % of MSP may have to be capped. Farmer can be incentivized to attempt higher price discovery. |
| **Assumptions** | The whole change relies on the assumption that farmers rather sell at lower price than MSP and there are traders who are willing to give a reasonable price which still gives immediate profit to farmers. |
| **External Risks** | Even the outlay for this program needs to be managed and any frauds can derail the whole program. |
| **Obstacles to Success** | The process for determining market prices needs to be credible as well as process to determine authenticity of quantum of produce and transactions. |
| **Knock-On Effects** | Farmers and traders may build healthy trusted relationships. |

**Table 12: Market Separations in Agriculture sector**

| Separation | Remarks |
|---|---|
| **Informational Separation** | Lack of information among farmers on prevailing price-bands in the market. |
| **Spatial Separation** | Distant from Mandis or market outlets and formal financial institutions may be a blocker |
| **Temporal Separation** | When a farmer wants to sell Government may not be buying thus forcing farmer to sell to aggregators. |
| **Financial Separation** | Inability to access institutional credit at lower interest rates. Inability to reinvest earning. |
| **Capability Separation** | Ability to shift to crops that have export potential. Ability to absorb risks in case of failure. |

Next, we look at what it takes to operationalize Tantra Framework to facilitate the analysis of Agricultural Sector in a holistic manner that is extensible to societal context. using Tantra Framework. Tantra Framework should address lack of high-quality information in pertaining to agricultural sector contextualized in rural economy and economy at large. Considering the size and complexity of agricultural sector, we have chosen the ITIL change management methodology of BMC, which is typically used in large and complex enterprises. See Table 13 below.

**Table 13: Operationalizing Agricultural Ecosystem using Tantra Framework**

| #. | Topic | Remarks |
|---|---|---|
| 1 | Issues in the current system | Lack of growth, reduced profitably, stagnant investment in agricultural sector along with continued poverty and sustained degradation of environment. |
| 2 | Change Agents | Government Departments, Farming communities and rural populace can act as change agents and participate in the change process/. |
| 3. | Desired Change | To promote agricultural sector as well as rural economy as engines of growth and spread prosperity among large sections of people. |
| 4 | List of Services to be built using Tantra Framework | A digital application/portal needs to be built that makes use of Tantra Framework to capture inputs as well as to share information. Framework enables collection and analysis of data in line with descriptive, normative and transformative view-points. |
| 5 | Strategic Alignment between Objectives and services | Having high quality information flow in an ongoing basis can enable sustained social change and progression towards prosperity. |
| 6 | Design of change Process with Purpose | Hence data collected is collected at sectoral, societal and economic levels. Processes need to be instituted to collect data from both formal and informal sources periodically. Data related to sectoral, social and economic indicators should be |

| | | assimilated in the Tantra Framework. This can help design better interventions that need to be attempted. |
|---|---|---|
| 7 | **Transition Process** | New processes should be instituted covering region after region where data is frequently collected and analyzed. They should interoperate with existing processes in a harmonious manner. |
| 8 | **Operational aspects and organizing actors/Institutions** | Existing Government organizations and farmer institutions can be asked to help out while taking care to maintain the sanctity of data collection process |
| 9 | **Process for collecting feedback** | Feedback from all stakeholders can be collected and shared as appropriate. Feedback should be used to achieve continuous service improvement. |
| 10 | **Training Plan** | The officials and people need to be trained with online material as well as off-line programs. |
| 11 | **Future Roadmap** | The scope can be enhanced to include crop analytics in the future. |

## 5. Transforming the Indian Agricultural Sector

To transform the Indian Agricultural Sector, it is critical to reduce the mutual dependence between the sector and the Government. In this section we explore an approach to achieve that. The role currently played by Government is depicted in Figure 3 below.

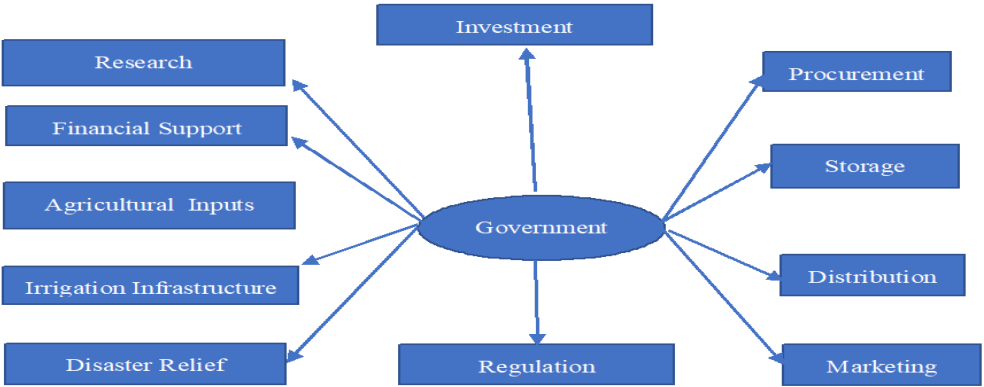

**Figure 3: Government's role in Indian Agricultural Sector**

Government of India invests in agriculture in variety of ways. Government invests in irrigation facilities, last mile connectivity of irrigated water to farmers, storage facilities in the form of Steel Silos, setting up markets for agricultural produce which go by the name APMC. Government has invested in Food Corporation of India, which is a huge nation-wide organization. Government has invested in agricultural research. Government regulates the agricultural sector heavily. Government disallows farmers from selling their goods except through authorized markets. Government disallows hoarding and cartelization. Government has export restrictions when prices go up. Undoubtedly there is need to do a retrospective on the foot-print of Government which has steadily expanded over the last seven decades.

This level of Government involvement is definitely not sustainable. On another notes agriculture itself should use sustainable practices via-a-vis environment. Considering both these, we look at the eco-system paradigm to design the required transformation in the agricultural sector. Peltoniemi and Vuori (2004) have described variety of ecosystems such as biological, industrial, economic, business, digital and social ecosystems in their paper.

In this paper we look at Agricultural Sector as a business ecosystem rather than natural eco-system which indeed it is. The key properties of Business ecosystem study are, coevolution, self-organization emergence and adaptation.

- **Coevolution** is the mutual change of species/organizations that interact with each other. Pagie(1999) discusses three types of coevolution, namely competitive, mutualistic and exploitative. *Competitive coevolution* occurs between species which are limited by the same resources. *Mutualistic coevolution*, on the other hand, comprises reciprocal relations where all the participants benefit from the interaction and change in the direction of better compatibility. This can also happen when organizations develop capabilities for cooperation and complementation in order to compete with a third party. Exploitative coevolution may be detected in a situation where an organization is significantly more powerful than the others. This could happen in the context of a large corporation and its suppliers.
- **Self-organization** in a business ecosystem context, implies the absence of a central or outside controller. Organizations make their own decisions based on imperfect, perhaps locally restricted, knowledge that they possess. Self-organization is enabled by the market economy system. In any real-life business ecosystem, however, there are interventions by the public sector, such as business subsidies, import duties and publicly funded development projects. These can be seen either as inhibiting self-organization or as creating enabling structures for self-organization.
- According to Smith and Stacey (1997) as **emergence** takes place, the behavior observed at the macro level is not obvious while examining the behavior at the micro level. "Emergence means that the links between individual agent actions and the long-term systemic outcome are unpredictable" They also state that emergence makes it impossible for one actor to control the whole system, whether inside or outside the system. Instead the behavior emerges Emergence is a phenomenon that arises from organization-level motives and actions that lead to unpredictable and even surprising population-level behavior. It is induced by each organization's restricted knowledge of its environment, of its options and of the outcomes of those options. Thus, the phenomenon may be amplified in the population and result in an unanticipated situation.
- The fourth property of business ecosystems is **adaptation.** According to Holland(1992,1995)adaptation generates "structures of progressively higher performance". Holland) suggests that there are three components associated to adaptation: the environment, the adaptive plan, and a measure of performance/fitness. Adaptation can be criticized for the passive role of environment. Adapting always means adapting to something, and it incorporates the thought that the adapting unit is not capable of having an effect on its environment. When the environment changes, a business ecosystem adapts to changed conditions by emergence, co-evolution and self-organization

As ecosystems expand and competition becomes harder, the businesses/organizations must develop "dynamic capabilities". The dynamic capabilities are signature processes/offerings which give an edge to a business, in contrast to ordinary capabilities which competitors can easily emulate. Next, we look at applying the concepts coevolution, self-organization, emergence and adaptation to Indian Agricultural Sector. The data assimilated through Tantra Framework can provide the foundation to architect the agricultural ecosystem, by drawing in participants and then observe/influence them. See Table 14 below.

**Table 14: Modelling Agriculture Sector as Business Ecosystem**

| |
|---|
| **Coevolution** |
| - Farmers should compete with each other instead of simply relying on Government. They also need to compete globally with best of breed practices.<br>- Under mutualistic coevolution Government nudges ecosystem players to seek balanced value capture among participants of value chain. The consumers should be educated to pay a bit more for food products as a fraction of their consumption basket pay-out. Aggregators, intermediaries and wholesalers should be nudged to make farming more lucrative by sharing greater portion of profits with farmers. Again, farmers should give fair pay to farm workers. Farm owners should pass on benefits to tenant farmers when Government routes the benefits through them.<br>- The contract farming companies as well as food processing companies should not operate in exploitative manner. |
| **Self-organization** |
| - Farmers should evolve flexible organization where they can mutually lease or pool land, labour, capital and produce so that they can collectively profit more. Co-operatives are successful in Dairy products.<br>- There should be greater adoption of crop insurance as well as allied businesses to agriculture. Here the farmer organizations can play a critical role.<br>- Traders can participate in E-NAMs and build trusted relationship across the nation.<br>- They can form advocacy groups which balance interest of farm owners, tenant farmers, farmer families and farm workers. They should be able to negotiate as a group with investors and lenders. For that they should spread better credit culture in farming community. |
| **Emergence** |
| Jean-François Ponge(2005) proposed 3 different models to explain emergence: bubble, wave and crystal. In the bubble model molecules are organized over the surface of a bubble giving that strength to the bubble. At the other extreme in the crystal model any part of the crystal has the same structure as the larger crystal. The intermediate model of waves is the most suitable to study emergence in Indian Agricultural Sector. Ponge, talks about traffic jams and how individual actions of vehicles make the whole traffic move. We need to improvise this model for agricultural sector where every vehicle that takes detour frees up the overall traffic. These detours may be adoption of ZBNF or organic farming which removes the need for fertilizer subsidies. Choice of drip irrigation to switch over to crops that need less water. From consumer side choosing Direct Benefit Transfer in place of food grains can be a change. From the standpoint of Government progressively loosening controls on trading and exports can be the right prescription. Overall, we need many individual/group/institutional actions which can lead to emergence of Indian Agriculture sector as thriving business ecosystem. Tantra Framework can help track the emergence in a fine-grained manner. |
| **Adaptation** |
| Adaptation is about adapting to environment whether it is natural, fiscal, social or political. Adopting sustainable farming practices is no longer a choice rather it is sheer necessity. In the same way relying on excessive financial support from Government may no longer work. |

## 6. Concluding remarks

India's agriculture sector is highly complex due to variety of factors. Nearly half-a-billion people are directly or involved with responsibility to feed a nation with billion-plus population. Government intervenes heavily in the sector. In addition, excessive use of fertilizers and pesticides has had harmful effect on environment. We have explained how Tantra Framework can be leveraged to represent sectoral information using descriptive, normative and transformative view-points. In conjunction with Tantra Framework we have made use the business ecosystem paradigm for Agricultural sector. We have elaborated how different ecosystem properties such as coevolution, self-organization, emergence and adaptation can be used to analyze and transform the agricultural sector. To gain from this approach, we need ability to collect humongous amounts of data which is possible only with cooperation from Government. Once we have that we need to strategize how to optimally leverage plethora of data Government already collects and lay out a new road map for agricultural data engineering. All this may require an overarching digital governance/information management architecture which Government supports, accepts and promotes. One good sign is that 2018-19 Economic Survey of India, alludes to the requirement for "Enterprise Architecture for Governance" where multiple databases are unified. Hopefully Tantra Framework can fulfill that requirement.